\documentclass[preprint, superscriptaddress, showpacs,preprintnumbers,amsmath,amssymb,prb]{revtex4}
\usepackage{graphicx}

\begin{document}

\thispagestyle{empty}

\title{Comparison of hydrodynamic model of graphene with
recent experiment on measuring the Casimir interaction}

\author{
G.~L.~Klimchitskaya}
\affiliation{Central Astronomical Observatory at Pulkovo of the Russian Academy of Sciences, St.Petersburg,
196140, Russia}
\affiliation{Institute of Physics, Nanotechnology and
Telecommunications, St.Petersburg State
Polytechnical University, St.Petersburg, 195251, Russia}

\author{
V.~M.~Mostepanenko}
\affiliation{Central Astronomical Observatory at Pulkovo of the Russian Academy of Sciences, St.Petersburg,
196140, Russia}
\affiliation{Institute of Physics, Nanotechnology and
Telecommunications, St.Petersburg State
Polytechnical University, St.Petersburg, 195251, Russia}

\begin{abstract}
We obtain the reflection coefficients from a graphene sheet
deposited on a material substrate under a condition that
graphene is described by the hydrodynamic model.
Using these coefficients, the gradient of the Casimir force
in the configuration of recent experiment is calculated in
the framework of the Lifshitz theory.
It is shown that the hydrodynamic model is excluded by the
measurement data at the 99\% confidence level over a wide
range of separations.
{}From the fact that the same data are in a very good
agreement with theoretical predictions of the Dirac model
of graphene, the low-energy character of the Casimir
interaction is confirmed.
\end{abstract}

\maketitle

\section{Introduction}

Graphene is a two-dimensional sheet of carbon atoms which attracts
much experimental and theoretical attention due to unique physical
properties and great promise for various applications.\cite{1}
The most often used approach describing the electric and optical
properties of graphene is the Dirac model.\cite{2}
It is applicable at low energies up to a few eV and assumes the
linear dispersion relation for massless quasiparticles, which move
with a Fermi velocity rather than with the speed of light.\cite{3}
The Dirac model was applied by many authors to calculate the
van der Waals, Casimir and Casimir-Polder interactions in layer
systems including graphene (see, for instance, Refs.\cite{4,5,6,7,8}).
The most straightforward formalism for performing these calculations
uses the polarization tensor in (2+1)-dimensional
space-time.\cite{9,10,11,12,13,14}
Recently an equivalence of the formalisms exploiting the polarization
tensor and the density-density correlation function has been
proven.\cite{15} Furthermore, the formalism of the polarization
tensor was applied\cite{16,17} for comparison with the measurement
data of recent experiment\cite{18} and demonstrated a very good
agreement.

Another approach used in theoretical description of the properties
of graphene is the hydrodynamic model.\cite{19,20}
This model considers graphene as an infinitesimally thin
positively charged sheet, carrying a homogeneous fluid with some
mass and negative charge densities. In the framework of this model,
the dispersion relation for quasiparticles in graphene is quadratic
with respect to the momentum. The hydrodynamic model was also
considered and
applied to calculate the van der Waals, Casimir and Casimir-Polder
interactions in many papers.\cite{11,12,21,22,23,23a,24,25a,25,25b}
It was found\cite{11} that in the interaction of graphene with
either a Si or a Au plate the hydrodynamic model predicts larger
magnitudes of the Casimir free energy than the Dirac model.

In this paper, we compare theoretical predictions of the hydrodynamic
model with the experimental data of recent measurement\cite{18}
 of the gradient
of the Casimir force between a Au-coated sphere and a graphene sheet
deposited on a SiO${}_2$ film covering a Si plate.
For this purpose, we derive exact expressions for the reflection
coefficients of the electromagnetic oscillations on a three-layer
structure, where one layer is a two-dimensional sheet described by the
hydrodynamic model, whereas two other layers are described by the
frequency-dependent dielectric permittivities.
Then, the Casimir force and its gradient are calculated by using the
standard Lifshitz theory\cite{26,27} in the proximity force
approximation\cite{27} (PFA). We demonstrate that theoretical
predictions of the hydrodynamic model are excluded by the measurement
data at the 99\% confidence level over a wide region of separations
between the sphere and the graphene sheet. This allows to conclude
that the hydrodynamic model of graphene does not describe such physical
phenomena, as the van der Waals and Casimir forces.

The paper is organized as follows. In Sec.~II we derive the reflection
coefficients for a graphene-coated substrate under an assumption that
graphene is described by the hydrodynamic model. Using these reflection
coefficients, in Sec.~III we calculate the gradient of the Casimir
force in the experimental configuration of recent experiment\cite{18}
and compare the theoretical results with the experimental data.
Section~IV contains our conclusions and discussion.

\section{Reflection coefficients in the hydrodynamic model}

We consider the amplitude reflection coefficients $R^{(g,s)}$ from
a graphene sheet deposited on a thick plate (semispace) made of an
ordinary material. Let us denote the reflection coefficient from
a freestanding graphene sheet by $r^{(g)}$ and from a semispace in
vacuum by $r^{(s)}$. Then, for the transverse magnetic (TM) and
transverse electric (TE) polarizations of the electromagnetic
field one obtains \cite{16}
\begin{equation}
R_{\rm TM,TE}^{(g,s)}=\frac{r_{\rm TM,TE}^{(g)}+
r_{\rm TM,TE}^{(s)}(1\mp 2r_{\rm TM,TE}^{(g)})}{1-
r_{\rm TM,TE}^{(g)}r_{\rm TM,TE}^{(s)}}.
\label{eq1}
\end{equation}
\noindent
Here, the signs minus and plus should be chosen for the TM and TE
polarizations, respectively.

In the framework of the hydrodynamic model, the reflection
coefficients $r^{(g)}$ for a graphene sheet in vacuum calculated
at the Matsubara frequencies along the imaginary frequency axis
take the form \cite{11,12,21,22,23,24,25,27}
\begin{eqnarray}
&&
r_{\rm TM}^{(g)}\equiv r_{\rm TM}^{(g)}(i\xi_l,k_{\bot})=
\frac{c^2q_lK}{c^2q_lK+\xi_l^2},
\nonumber \\
&&
r_{\rm TE}^{(g)}\equiv r_{\rm TE}^{(g)}(i\xi_l,k_{\bot})=
-\frac{K}{K+q_l}.
\label{eq2}
\end {eqnarray}
\noindent
Here, the Matsubara frequencies are $\xi_l=2\pi k_BTl/\hbar$,
$k_B$ is the Boltzmann constant, $T$ is the temperature,
$l=0,\,1,\,2,\,\ldots\,$, $q_l^2=k_{\bot}^2+\xi_l^2/c^2$,
$k=|\mbox{\boldmath$k$}_{\bot}|$, and $\mbox{\boldmath$k$}_{\bot}$
is the projection of the wave vector on the plane of graphene.
The quantity $K$ is the the single parameter characterizing graphene in the
framework of the hydrodynamic model. It has the meaning of the wave
number of a graphene sheet and is determined by the parameters of the
hexagonal structure of graphite (one $\pi$-electron per atom, resulting
in two $\pi$-electrons per hexagonal cell). Calculation leads to
 \cite{19,20,21}
\begin{equation}
K=2\pi\frac{ne^2}{mc^2}=6.75\times 10^5\,\mbox{m}^{-1},
\label{eq3}
\end{equation}
\noindent
where $e$ and $m$ are the charge and mass of $\pi$-electrons
and $n=4/(3\sqrt{3}l^2)$ with $l=1.421\,${\AA} being the side
length of the hexagon in a crystal lattice. The wave number
in Eq.~(\ref{eq3}) corresponds to the frequency
$\omega_K=cK=2.02\times 10^{14}\,$rad/s.

The reflection coefficients $r^{(s)}$ from the boundary plane
of a semispace described by the dielectric permittivity
$\varepsilon_{1l}\equiv\varepsilon_1(i\xi_l)$ are the well
known Fresnel coefficients
\begin{eqnarray}
&&
r_{\rm TM}^{(s)}\equiv r_{\rm TM}^{(s)}(i\xi_l,k_{\bot})=
\frac{\varepsilon_{1l}q_l-k_{1l}}{\varepsilon_{1l}q_l+k_{1l}},
\nonumber \\
&&
r_{\rm TE}^{(s)}\equiv r_{\rm TE}^{(s)}(i\xi_l,k_{\bot})=
\frac{q_l-k_{1l}}{q_l+k_{1l}},
\label{eq4}
\end {eqnarray}
\noindent
where
\begin{equation}
k_{1l}^2\equiv k_1^2(i\xi_l,k_{\bot})=k_{\bot}^2+
\varepsilon_{1l}\frac{\xi_l^2}{c^2}.
\label{eq5}
\end{equation}
\noindent
Substituting Eqs.~(\ref{eq2}) and (\ref{eq4}) in Eq.~(\ref{eq1}),
we find the reflection coefficients from a graphene sheet deposited on
a semispace made of ordinary material
\begin{eqnarray}
&&
R_{\rm TM}^{(g,s)}\equiv R_{\rm TM}^{(g,s)}(i\xi_l,k_{\bot})=
\frac{\varepsilon_{1l}q_l\xi_l^2-k_{1l}\xi_l^2+
2c^2q_lKk_{1l}}{\varepsilon_{1l}q_l\xi_l^2+k_{1l}\xi_l^2+
2c^2q_lKk_{1l}},
\nonumber \\[2mm]
&&
R_{\rm TE}^{(g,s)}\equiv R_{\rm TE}^{(g,s)}(i\xi_l,k_{\bot})=
\frac{q_l-k_{1l}-2K}{q_l+k_{1l}+2K}.
\label{eq6}
\end {eqnarray}

For computational purposes, we express the reflection coefficients
(\ref{eq6}) in terms of the dimensionless variables
\begin{equation}
y=2aq_l,\qquad \zeta_l=\frac{2a\xi_l}{c},
\label{eq7}
\end{equation}
\noindent
where $a$ is a parameter having the dimension of length
(in the next section $a$ has the meaning of separation distance
between a graphene-coated substrate and a sphere).
Then one arrives at
\begin{eqnarray}
&&
R_{\rm TM}^{(g,s)}\equiv R_{\rm TM}^{(g,s)}(i\zeta_l,y)=
\frac{\varepsilon_{1l}y\zeta_l^2-\tilde{k}_{1l}\zeta_l^2+
2\tilde{K}y\tilde{k}_{1l}}{\varepsilon_{1l}y\zeta_l^2+
\tilde{k}_{1l}\zeta_l^2+2\tilde{K}y\tilde{k}_{1l}},
\nonumber \\[2mm]
&&
R_{\rm TE}^{(g,s)}\equiv R_{\rm TE}^{(g,s)}(i\zeta_l,y)=
\frac{y-\tilde{k}_{1l}-2\tilde{K}}{y+
\tilde{k}_{1l}+2\tilde{K}},
\label{eq8}
\end {eqnarray}
\noindent
where
\begin{equation}
\tilde{k}_{1l}^2=4a^2k_{1l}^2=y^2+(\varepsilon_{1l}-1)
\zeta_l^2,
\qquad
\tilde{K}=2aK.
\label{eq9}
\end{equation}

\section{Comparison of the hydrodynamic model with the
measurement data}

In the first experiment on the Casimir effect in systems
including graphene,
the gradient of the Casimir force was measured between a Au-coated
hollow glass sphere of radius $R=54.1\,\mu$m and a graphene sheet
deposited on a SiO${}_2$ film covering a Si plate.\cite{18}
The thickness of a SiO${}_2$ film  was $D=300\,$nm.
The thickness of a Si plate ($500\,\mu$m) was large enough to
consider it as a Si semispace when calculating the Casimir force.
In a similar way, the Au coating on a sphere resulted in the same
Casimir force as an all-Au sphere. Measurements were performed by
means of dynamic atomic microscope operated in the frequency-shift
technique.\cite{28,29,30,31,32}
The force-distance relations were obtained with different applied
voltages (20 repetitions) and with applied compensating voltages
(22 repetitions) over the separation region from 224 to 500\,nm
for two different graphene samples. All the mean gradients of the
Casimir force were found to be  in a very good mutual agreement in
the limits of the experimental errors.\cite{18}
As an example, in Fig.~1(a,b) typical mean gradients of the Casimir
force (the first sample, the measurement results obtained with applied
compensating voltage) are shown as crosses at different separations
$a$ between the sphere and the plate. The vertical arms of the crosses
indicate twice the total error $\Delta F^{\prime}=0.64\,\mu$N/m in
measurements of the gradient of the Casimir force, and the horizontal
arms are twice the error $\Delta a=0.4\,$nm in measurement of absolute
separations. These errors were found at the 67\% confidence level,
i.e., the true values of the force gradients and separations with a
probability of 67\% belong to the intervals
$[ F^{\prime}(a)-\Delta F^{\prime}, F^{\prime}(a)+\Delta F^{\prime}]$ and
$[a-\Delta a,a+\Delta a]$, respectively.

Using the Lifshitz theory and the PFA, the gradient of the Casimir
force between a Au sphere and a graphene sheet
deposited on a SiO${}_2$ film covering a Si plate (semispace) is
given by
\begin{eqnarray}
&&
F^{\prime}(a)=\frac{k_BTR}{4a^3}\sum_{l=0}^{\infty}\!
{\vphantom{\sum}}^{\prime}\int_{\zeta_l}^{\infty}\!\!\! y^2dy
\left[\frac{r_{\rm TM}^{(\rm Au)}(i\zeta_l,y)
R_{\rm TM}^{(g,f,s)}(i\zeta_l,y)}{e^y-
r_{\rm TM}^{(\rm Au)}(i\zeta_l,y)
R_{\rm TM}^{(g,f,s)}(i\zeta_l,y)}\right.
\nonumber \\
&&~~~~
\left.+
\frac{r_{\rm TE}^{(\rm Au)}(i\zeta_l,y)
R_{\rm TE}^{(g,f,s)}(i\zeta_l,y)}{e^y-
r_{\rm TE}^{(\rm Au)}(i\zeta_l,y)
R_{\rm TE}^{(g,f,s)}(i\zeta_l,y)}\right],
\label{eq10}
\end{eqnarray}
\noindent
Here, $T=300\,$K is the temperature at the laboratory, the prime on the
summation sign multiplies by 1/2 the term with $l=0$, and the dimensionless
variables $y$ and $\zeta_l$ are introduced in Eq.~(\ref{eq7}).
The reflection coefficients from a Au semispace (which replaces a sphere
in the PFA) are given by Eq.~(\ref{eq4}). In terms of dimensionless
variables, they take the form
\begin{eqnarray}
&&
r_{\rm TM}^{(\rm Au)}(i\zeta_l,y)=
\frac{\varepsilon_l^{\,(\rm Au)}y-\tilde{k}_l^{(\rm Au)}
}{\varepsilon_l^{\,(\rm Au)}
y+\tilde{k}_l^{(\rm Au)}},
\nonumber \\
&&
r_{\rm TE}^{(\rm Au)}(i\zeta_l,y)=
\frac{y-\tilde{k}_l^{(\rm Au)}}{y+
\tilde{k}_l^{(\rm Au)}},
\label{eq11}
\end{eqnarray}
\noindent
where, in accordance to Eq.~(\ref{eq9}),
\begin{equation}
\tilde{k}_l^{(\rm Au)}=
[y^2+(\varepsilon_l^{\,(\rm Au)}-1)\zeta_l^2]^{1/2}.
\label{eq12}
\end{equation}
\noindent
The reflection coefficients from a graphene sheet
deposited on a SiO${}_2$ (fused silica) film covering a Si plate
are expressed by the standard formulas of the Lifshitz theory
between layered structures\cite{27,33}
\begin{equation}
R_{\rm TM,TE}^{(g,f,s)}(i\zeta_l,y)=
\frac{R_{\rm TM,TE}^{(g,s)}(i\zeta_l,y)+
r_{\rm TM,TE}^{(f,s)}(i\zeta_l,y)e^{-2D\tilde{k}_{1l}/(2a)}}{1+
R_{\rm TM,TE}^{(g,s)}(i\zeta_l,y)
r_{\rm TM,TE}^{(f,s)}(i\zeta_l,y)e^{-2D\tilde{k}_{1l}/(2a)}}.
\label{eq13}
\end{equation}
\noindent
Here, the reflection coefficients $R_{\rm TM,TE}^{(g,s)}$ describe
the reflection from a graphene sheet deposited on a SiO${}_2$
semispace. If graphene is described by the hydrodynamic model,
they are given by Eq.~(\ref{eq8}) with
$\varepsilon_{1l}=\varepsilon^{({\rm SiO}_2)}(ic\zeta_l/2a)$.
The coefficients $r_{\rm TM,TE}^{(f,s)}$ describe the reflection
on the boundary plane between the two semispaces made of SiO${}_2$
and Si. They are the standard Fresnel reflection coefficients:
\begin{eqnarray}
&&
r_{\rm TM}^{(f,s)}(i\zeta_l,y)=
\frac{\varepsilon_{2l}\tilde{k}_{1l}-
\varepsilon_{1l}\tilde{k}_{2l}}{\varepsilon_{2l}\tilde{k}_{1l}+
\varepsilon_{1l}\tilde{k}_{2l}}
\nonumber \\
&&
r_{\rm TE}^{(f,s)}(i\zeta_l,y)=
\frac{\tilde{k}_{1l}-\tilde{k}_{2l}}{\tilde{k}_{1l}+\tilde{k}_{2l}},
\label{eq14}
\end{eqnarray}
\noindent
where
$\varepsilon_{2l}\equiv\varepsilon^{\,(\rm Si)}(ic\zeta_l/2a)$
and $\tilde{k}_{2l}$ is defined similar to Eq.~(\ref{eq9}) with
a replacement of  $\varepsilon_{1l}$ with $\varepsilon_{2l}$.

The quantity $\varepsilon_{l}^{(\rm Au)}\equiv\varepsilon^{(\rm Au)}(ic\zeta_l/2a)$
entering Eq.~(\ref{eq11}) is found \cite{27,28} using the Kramers-Kronig
relation from the measured optical data \cite{34} for
${\rm Im}\,\varepsilon^{(\rm Au)}$ extrapolated to zero frequency either by the
Drude model with the plasma frequency $\omega_p=9.0\,$eV and relaxation
parameter $\gamma=0.035\,$eV or by the nondissipative plasma model.
Note that the above values of the Drude parameters are in a very good
agreement with the measured optical data.\cite{35}
Contrary to the expectations, the most precise experiments on measuring
the Casimir interaction between metallic surfaces \cite{27,28,29,30,31,36,37,38,39}
are in agreement with the theoretical predictions using the plasma model
extrapolation of the optical data and exclude the theoretical results using
the Drude model extrapolation.
Deep physical reasons why the plasma model extrapolation of the optical
data is in agreement with the most precise measurements and the Drude
model extrapolation is excluded by them remain unknown.
 Here, we perform all computations using both
extrapolations for Au. We find that for a sphere interacting with a graphene-coated
substrate, where graphene is described by the hydrodynamic model, the difference
arising from using different extrapolations is rather small. This allows to include it
in the magnitude of the theoretical error like it was done in the
case of  metal-graphene
interaction computed using the Dirac model of graphene.\cite{11,16,18}

In order to calculate the reflection coefficients (\ref{eq13}) one also needs
the values of $\varepsilon^{(\rm Si)}$ and  $\varepsilon^{({\rm SiO}_2)}$
at the imaginary Matsubara frequencies. The B-doped Si plate used in the
experiment \cite{18} had a resistivity between 0.001 and 0.005\,$\Omega$\,cm.
This corresponds \cite{40} to a charge carrier density between
$1.6\times 10^{19}$ and $7.8\times 10^{19}\,\mbox{cm}^{-3}$, i.e., well above
the critical density at which the dielectric-to-metal phase transition
accurs.\cite{41} Then one obtains for the plasma frequency \cite{42} the
values between $5\times 10^{14}$ and $11\times 10^{14}\,$rad/s and for the
relaxation parameter \cite{31} $\gamma\approx 1.1\times 10^{14}\,$rad/s.
These Drude parameters were used to extrapolate the optical data \cite{43}
for ${\rm Im}\,\varepsilon^{(\rm Si)}$ to zero frequency by means of either
the Drude or the plasma model. Finally, the dielectric permittivity of Si
at the imaginary Matsubara frequencies was found by means of the Kramers-Kronig
relation like this was done previously in the literature.\cite{44}
Different types of extrapolation for a Si lead to only minor differences in
the computed force gradients in the experimental configuration.
This is also taken into account in the theoretical error. For the dielectric
permittivity of SiO${}_2$ an accurate analytic expression \cite{45} has
been used.

The theoretical force gradients using the hydrodynamic model of graphene were
computed by Eqs.~(\ref{eq8}), (\ref{eq10}), (\ref{eq11}), (\ref{eq13}), and
(\ref{eq14}). The computational results were corrected for the presence of
surface roughness which contribution does not exceed 0.1\% in this experiment.\cite{16}
The computed gradients of the Casimir force are shown as blue bands in
Fig.~1(a,b) over the entire measurement range. The uncertainty in the values
of $\omega_p$ of Si and the differences between the predictions of the
Drude and plasma model extrapolations of the optical data for Au and Si
determine the theoretical error, which is taken into account in the width of
the bands. As is seen in Fig.~1, the theoretical description of graphene
using the hydrodynamic model is excluded by the data at the 67\% confidence level over
the entire measurement range from 224 to 500\,nm.
With increasing confidence level, the error bars of the mean force gradients and separation
distances are also increasing. Thus, at the 95\% confidence level the maximum increase
of the error bars is by a factor of two.\cite{27,46}
Note that if the errors are determined at the 95\% or 99\% confidence
levels the true values of the force gradients and separations belong
to the wider intervals
$[ F^{\prime}(a)-\Delta F^{\prime}, F^{\prime}(a)+\Delta F^{\prime}]$ and
$[a-\Delta a,a+\Delta a]$ with probabilities of 95\% and
99\%, respectively.
Taking this into account,
it can be seen that over the range of separations from 224 to 450\,nm
theoretical predictions of the hydrodynamic model are excluded by the data
at a higher, 95\% confidence level. If we further increase the confidence level
up to 99\%, it is easily seen that
theoretical predictions of the hydrodynamic model are still excluded,
but this time over a more narrow separation range up to 360\,nm.

It is interesting also to compare the theoretical predictions of the hydrodynamic
model with theoretical predictions of the Dirac model over a wider separation
region from 220\,nm to $1\,\mu$m. In Fig.~2 the gradients of the Casimir force
in the configuration of an experiment\cite{18} computed in this paper using the
hydrodynamic model (the dashed line) and using the Dirac model\cite{16,17}
(the solid line) are shown as functions of separation. As is seen in Fig.~2,
the predictions of the hydrodynamic model remain to be larger than the
experimentally consistent predictions of the Dirac model.
The physical reason why the hydrodynamic model is not suitable for
theoretical description of the Casimir force in layered systems
including graphene may be in the linear dispersion relation inherent
to graphene at low energies. This property makes a big difference
between graphene and all types of ordinary dielectrics and metals.

\section{Conclusions and discussion}

In this paper, we have compared the measurement results for the gradient of the
Casimir force between a Au-coated sphere and a graphene-coated substrate \cite{18}
with theoretical predictions of the hydrodynamic model of graphene used by many
authors in previous literature. For this purpose, the reflection coefficients
from the three-layer structure, where the first layer is graphene described by
the hydrodynamic model and two other layers are described by the
frequency-dependent dielectric permittivities, have been obtained.
It was shown that the hydrodynamic model of graphene is excluded by the
measurement data over the entire measurement range from 224 to 500\,nm
at the 67\% confidence level.
Over a narrower separation region from 224 to 360\,nm an exclusion of
the hydrodynamic model by the data at even higher 99\%
confidence level is demonstrated.

The same experimental data \cite{18} was recently shown \cite{16} to be  in a very
good agreement with theoretical predictions using the Dirac model of graphene.
Keeping in mind that the Dirac model is applicable at energies below a few eV,
the results of this paper provide additional arguments in favor of the low-energy
character of the Casimir interaction. In future it would be interesting to apply
the hydrodynamic model for theoretical description of the reflectivity
 of graphene at higher energies, outside the applicability region
of the Dirac model, and perform a comparison with respective experimental
results.


\begin{figure}[b]
\vspace*{-6cm}
\centerline{\hspace*{1cm}
\includegraphics{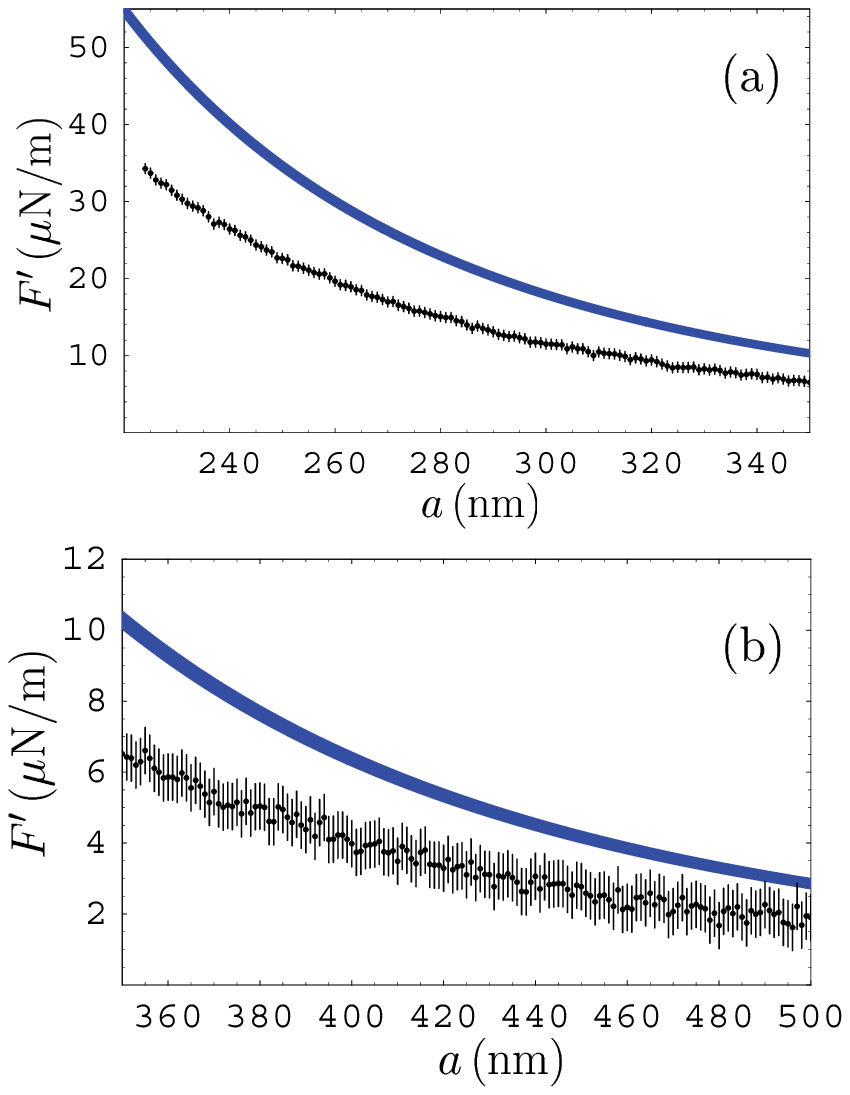}
}
\vspace*{-10cm}
\caption{\label{fg1}(Color online)
The experimental data for
the gradient of the Casimir force
between a Au-coated sphere and a graphene sheet
deposited on a
SiO${}_2$ film covering a Si plate (the first sample)
are shown as crosses plotted at a 67\% confidence level
over different separation regions
(a) from 224 to 350\,nm and (b) from 350 to 500\,nm.
The respective theoretical predictions of the
hydrodynamic model of graphene are indicated as
the blue bands.
}
\end{figure}
\begin{figure}[b]
\vspace*{-12cm}
\centerline{\hspace*{1cm}
\includegraphics{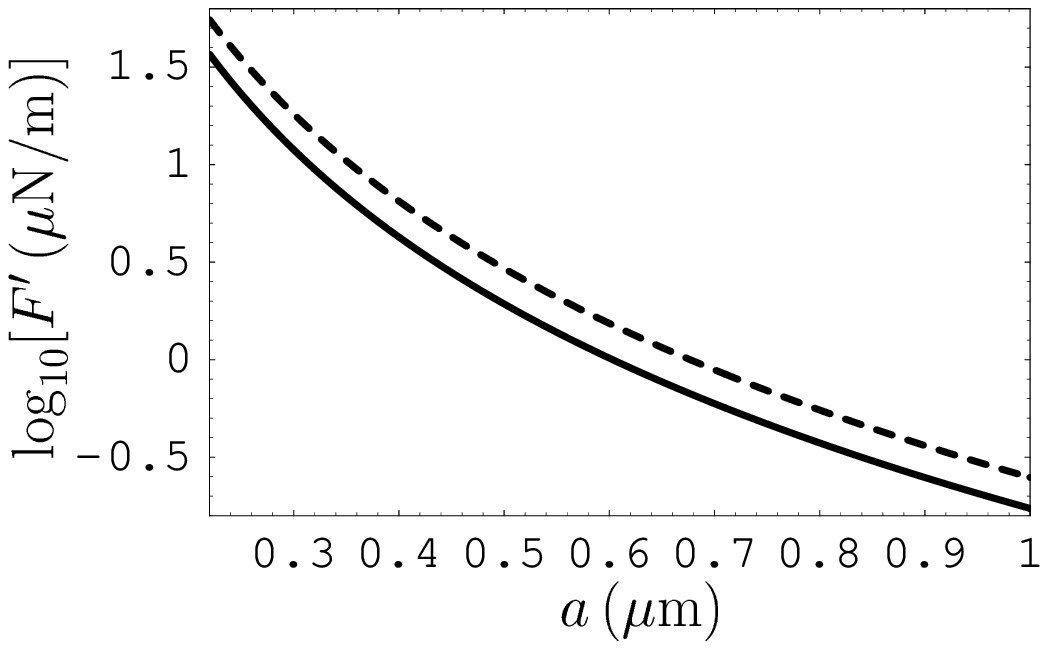}
}
\vspace*{-5cm}
\caption{\label{fg2}
The gradients of the Casimir force
between a Au-coated sphere and a graphene sheet
deposited on a
SiO${}_2$ film covering a Si plate
are computed using the
hydrodynamic model (the dashed line) and the
Dirac model (the solid line) as functions
of separation.
}
\end{figure}
\end{document}